\documentclass[a4paper,11pt]{article}
\usepackage{jcappub}  
\usepackage{graphicx}

\title{Self-consistent model of extragalactic neutrino flux from evolving blazar population}

\author[a,b]{A.Neronov} 
\author[a,c]{D.Semikoz}
\affiliation[a]{APC, Universit\'e Paris Diderot, CNRS/IN2P3, CEA/IRFU,
Observatoire de Paris, Sorbonne Paris Cit\'e, 119 75205 Paris, France}
\affiliation[b]{Astronomy Department, University of Geneva, Ch. d’Ecogia 16, Versoix, 1290, Switzerland}
\affiliation[c]{National Research Nuclear University MEPHI (Moscow Engineering 
Physics Institute), Kashirskoe highway 31, 115409 Moscow, Russia}
\abstract{
We study constraints on the population of neutrino emitting blazars imposed by the absence of doublets in astrophysical muon neutrino signal  and $z\simeq 0.3$ redshift of nearest identified neutrino-emitting blazar (an order of magnitude further away than the nearest $\gamma$-ray emitting blazar). We show that in spite of the absence of correlation of neutrino arrival directions with positions of gamma-ray emitting blazars, cumulative blazar flux could explain most of astrophysical neutrino flux measured in muon neutrino channel. This is possible if the population of neutrino emitting blazars has experienced rapid positive evolution at least as $(1+z)^5$  at $z\lesssim 1$. Such a model avoids previously derived constraint on the low level of blazar contribution to extragalactic neutrino flux because gamma-ray and neutrino fluxes are dominated by different sets of blazars. Rapid evolution of neutrino emitting blazars could be explained by the fact that only high luminosity blazars hosting radiatively efficient accretion flows are efficient neutrino sources. }
\emailAdd{andrii.neronov@unige.ch}      
\emailAdd{dmitri.semikoz@apc.univ-paris7.fr}
\begin{document}

\maketitle

\section{Introduction}

Five years after discovery \cite{icecube_discovery}, the origin of astrophysical neutrino signal detected in "high-energy starting events" (HESE) \cite{hese} and muon neutrino \cite{muon} channels by IceCube telescope remains uncertain. 

The overall flux and spectral slope of the HESE signal  are consistent with the high-energy extrapolation of the gamma-ray flux detected by Fermi telescope up to the TeV band \cite{neronov_semikoz,neronov_semikoz_muon,neronov_semikoz_evidence,neronov_semikoz_lhc,neronov_semikoz_galactic1}.  Given that the TeV gamma-ray flux from the sky is dominated by the emission from the Milky Way, the steep-spectrum HESE neutrino flux could well be of Galactic origin. Anisotropy pattern of the signal does not reveal strong excess toward the Galactic Plane of the type predicted by models of cosmic ray propagation in the Milky Way disk \cite{galactic_icecube,galactic_icecube_antares,galactic_antares}. However, the template of Galactic neutrino signal in 100 TeV range is difficult to work out because those neutrinos are prodiuced by cosmic rays with energies in 1-10 PeV range which do not  propagate through the interstellar medium in the same way as 10 GeV cosmic rays which produce GeV gamma-ray emission \cite{giacinti18,neronov_semikoz}. 

Neutrino signal at energies higher than several hundred TeV sampled from the Northern hemisphere with muon neutrinos reveals harder spectrum compared to that of the HESE neutrino flux \cite{muon,icecube_combined}. This hardening could be due to the presence of extragalactic component of the astrophysical neutrino flux. The overall flux of the hard component is at the level consistent with the observed ultra-high-energy cosmic ray (UHECR) flux \cite{wb,giacinti15,kalashev17}. Cosmic ray acceleration possibly to UHECR energies in  extragalactic astronomical sources  is inevitably accompanied by interactions of freshly accelerated protons and atomic nuclei with matter and radiation environments in the source. These interactions produce neutrinos through decays of pions and neutrino flux from sources of UHECR is generically expected \cite{berezinsky69,berezinsky75,wb,giacinti15,kalashev17}. 

Radio-loud Active Galactic Nuclei (AGN) are among astronomical source classes in which physical conditions which enable acceleration of protons and nuclei to energies up to UHECR range are realised \cite{neronov_semikoz02,neronov_sibiryakov,njp}. Neutrino emission from AGN, and in particular from blazars, is widely discussed in the context of hadronic models of AGN activity \cite{mb89,mannheim,neronov_semikoz02,neronov_sibiryakov,tchernin,cerutti18,cerutti18_1,taylor18,tavecchio18,padovani18,txs_xray,txs_onemore}. Within hadronic model framework, neutrino emission is generically expected to be accompanied by GeV gamma-ray emission produced in result of development of electromagnetic cascade inside the neutrino emitting source. 

In this respect, it is surprising that brightest and/or nearest gamma-ray blazars do not appear as brightest neutrino sources \cite{neronov_semikoz_blazars,icecube_blazars}. Analysis by IceCube collaboration \cite{icecube_blazars} concludes that blazars dominating the gamma-ray sky observed by Fermi Large Area Telescope (LAT)  could not explain the observed level of the astrophysical neutrino flux. Analysis of Ref. \cite{neronov_semikoz_blazars} shows that hadronic emission does not dominate the energy output of blazars.  Nevertheless, the only extragalactic source for which evidence for neutrino signal was found, TXS 0506+056 \cite{0506_multiwavelength,0506_icecube}, is a blazar. 

In what follows we show that blazars could in fact provide significant contribution to the hard-spectrum astrophysical muon neutrino flux, once possible difference in the population of gamma-ray and neutrino emitting blazars is properly taken into account. As it is discussed in Ref. \cite{neronov_semikoz02} not all blazars are expected to be "neutrino-loud". Differences in overall gamma-ray and neutrino emission power are generically expected  because neutrinos are efficiently produced only in the presence of dense matter and radiation backgrounds \cite{mb89,mannheim,neronov_semikoz02,neronov_sibiryakov,tavecchio18}.  

We explore constraints on neutrino emitting blazar population imposed by observational properties of the neutrino signal: the absence of event clustering in neutrino arrival directions, a problem first noticed in the analysis of Ref. \cite{waxman}, and the fact that nearby blazars are not strong neutrino sources, with the nearest identified neutrino emitted blazar TXS 0506+056  at redshift $\simeq 0.3$ \cite{0506_redshift} which is by a factor $\sim 10$ further away than the closest gamma-ray blazar (Mrk 421). We show that these facts suggest that neutrino emitting blazars have experienced rapid cosmological evolution at recent epoch $z\lesssim 1$.  This provides explanation for the fact that nearby blazars which provide dominant contribution to extragalactic $\gamma$-ray flux do not provide dominant contribution to the extragalactic neutrino flux. 

\section{Clustering of neutrino arrival directions and nearest detectable source distance from Monte-Carlo simulations}

We re-asses the constraint on the properties of neutrino sources imposed by non-observation of clustering of neutrino arrival directions on the sky, first considered in Ref. \cite{waxman}, while adding an addiitonal constraint that the nearest representative of the blazar population which has yielded individually detectable neutrino signal  is at $z\simeq 0.3$. 

We consider "standard candle" type sources with luminosity function $\rho(L_E,z) dL_E$ (the comoving number density of sources  at a given redshift $z$ having spectral luminosities $L_E$ to $L_E+dL_E$ at an energy $E$) which is proportional to a $\delta$ function 
\begin{equation}
\label{eq:sc}
\rho(L_E,z)=\rho_*(1+z)^\zeta \delta(L_E-L_{E*}(E))
\end{equation} 
where $L_{E*}(E)\propto E^{1-\gamma}$ is assumed to be a powerlaw with the differential spectrum slope $\gamma$. We allow the standard candle luminosity to evolve with redshift as $(1+z)^\zeta$. Models considered in Monte-Carlo simulations described below assume either positive evolution  up to $z_*$ followed by no-evolution period between $z_*$ and $z_{max}=3$. Such evolution patterns are characteristic for blazar populations: flat spectrum radio quasars (FSRQ) \cite{ajello09_fsrq} and BL Lacs  \cite{ajello09_bllac} as well as to the parent populations of FSRQs and BL Lacs, Fanaroff-Riley radio galaxies of type I and II \cite{fri,frii} and to X-ray selected AGN \cite{hasinger05}. 

Assuming that neutrino flux is emitted into a jet with an opening angle $\theta_{jet}$ one could find that the average number of neutrino events detectable with a telescope with effective collection area $A_{eff}$ within exposure time $T_{exp}$ is related to the  luminosity $L_E$ as
\begin{eqnarray}
\label{eq:le}
N_\nu(E_*)&=&(1+z)^{2-\gamma}\frac{A_{eff}T_{exp}L_E(E_*)}{\pi\theta_{jet}^2 d_L^2}
\end{eqnarray}
where $d_L$ is the luminosity distance.

If the jet directions are randomly distributed, the probability to find a given source with a jet pointing in the direction of an observer is 
\begin{equation}
p_{obs}=\frac{\theta_{jet}^2}{2}
\end{equation}
so that the "effective" density of sources visible for an observer is 
\begin{equation}
\label{rhoeff}
\rho_{eff}(L_E,z)=\frac{\theta_{jet}^2\rho(L_E,z)}{2}
\end{equation}

The number of observable sources    in the redshift range $z$ to $z+dz$ and producing given number of neutrino events between $N_\nu$ and $N_\nu+dN_\nu$ is 
\begin{equation}
\label{eq:eta}
\eta(N_\nu,z) dN_\nu dz=\rho_{eff}(L_E,z)dL_E dV_C
\end{equation}
where $L_E$ is expressed through $N_\nu$ using eq. (\ref{eq:le}), $dV_C$ is the comoving volume element per steradian of the telescope field-of-view
\begin{equation}
dV_C=\frac{d_C^2}{H_0E(z)}dz
\end{equation}
where
\begin{equation}
E(z)=\sqrt{\Omega_{0,m}(1+z)^3+\Omega_{0,\Lambda}}
\end{equation}
with $\Omega_{0,m}, \Omega_{0,\Lambda}$ being the present day dark matter and dark energy  density parameters and $d_C$ is the comoving distance related to $d_L$ as $d_C=d_L/(1+z)$,
\begin{equation}
d_C=\frac{1}{H_0}\int_0^z\frac{dz'}{E(z')}
\end{equation}

Calculating  $dL_E$ from (\ref{eq:le}) and substituting the expression for the comoving volume element in the right hand side of Eq. (\ref{eq:eta}) gives
\begin{eqnarray}
&&\eta(N_\nu,z)  =\frac{\pi\theta_{jet}^4 d_L^4\rho(L_E,z) }{2H_0(1+z)^{4-\gamma}A_{eff}T_{exp}E(z)}\nonumber
\end{eqnarray}
integrating over redshifts gives the differential source count (the number of sources contributing between $N_\nu$ and $N_\nu+dN_\nu$ counts):
\begin{eqnarray}
&& n(N_\nu)=\int_0^\infty \eta(N_\nu,z)dz=\nonumber\\
&&\frac{\pi\theta_{jet}^4}{2H_0A_{eff}T_{exp}}\int_0^\infty\frac{d_L^4\rho(L_E,z)}{(1+z)^{4-\gamma}E(z)}dz
\end{eqnarray}
The total number of sources producing at least $m$ events within a given exposure is
\begin{equation}
{\cal N}_s(N_\nu>m)=\int_m^\infty n(N_\nu)dN_\nu
\end{equation}

The above expression does not take into account Poisson statistics of the signal from individual sources, which is important at low $m$ values.  To properly deal with small $m$ case, we use Monte-Carlo simulations of the signal from a source population. 

In our Monte-Carlo simulations we first  generate source distribution which we assume to be uniform throughout the comoving volume. For each source we ascribe a fixed luminosity depending on the source distance / redshift (in this sense, we assume "pure luminosity" evolution model, rather than "luminosity dependent density evolution" model which better suits the description of population of blazars \cite{ajello09_bllac,ajello09_fsrq}). Fixing the position and luminosity of each source, we calculate its expected relative contribution to the neutrino flux at Earth as a function of (properly redshifted) neutrino energy, assuming that all sources have powerlaw type spectra with the slope $\gamma=2$. Our calculation takes into account the source position on the sky, and the declination dependence of the IceCube effective area $A_{eff}(E)$ \cite{icecube_combined,muon}.

Next, we simulate the neutrino signal with total statistics $N_{\nu,tot}\simeq 24$ events from the simulated source population. This signal statistics corresponds to that of the published IceCube sample of muon neutrinos \cite{muon} with muon energy proxies above 200~TeV, if the residual atmospheric neutrino background (approximately one third of the muon neutrino sample) is removed. 

Comparing energy distribution of detected muons with that of the detected neutrinos (as estimated from Monte-Carlo simulations in Ref. \cite{muon}) we note that the two distributions repeat each other with a shift downward in energy by factor $\epsilon\simeq 0.1$. This suggests an estimate of muon energies in our simplified Monte-Carlo simulations. We assume that detected muons have experienced an order-of-magnitude energy loss before entering the IceCube detector. We retain only muons which arrive at the detector with energies above 200 TeV.  

Different sources from the simulated blazar source set contribute to the samples of $N_{\nu,tot}$ neutrino events proportionally to their relative  contribution to the overall neutrino flux on average. However, the Poisson nature of the low statistics neutrino signal leads to significant fluctuations of the relative source contributions to the signal. Sources which on average are expected to give one or less neutrino could occasionally produce doublets in the simulated signal. Sources which are on average expected to produce multiplets in the simulated neutrino signal could occasionally have a down-fluctuation and yield one or even zero contribution to the signal.

\section{Results }

\subsection{Non-evolving sources}

The number of sources (\ref{eq:sc}) which produce at least $m$ events within a given exposure  is 
\begin{eqnarray}
{\cal N}_s(N_\nu>m)=
\frac{\rho_{*,eff}}{H_0} \int_0^{z_m} \frac{d_L^2(1+z)^{\zeta-2} }{E(z)} dz
\end{eqnarray}
where  $z_m$ is the redshift at which the standard candle source produces on average $m$ events within a given exposure. This expression simply states that to find the number of sources producing more than $m$ counts in the telescope one has to count all the sources with jets pointing to the observer up to the distance at which a source with the luminosity $L_{E*}$  produces  $m$ counts on average.  

Non-observation of sources producing multiplet events in IceCube indicates that the effective source density is low enough so that typically there are no sources contained within a sphere of the radius at which an individual source produces one event.  Assuming that $z_m\ll 1$ one could find that sources produce on average $m$ events as a distance 
\begin{equation}
d_m= \sqrt{\frac{A_{eff}T_{exp}L_{E*}}{\pi\theta_{jet}^2m}}
\end{equation}
The condition that there is less than one source within the volume of the sphere with radius $d_m$ is then 
\begin{equation}
\rho_{*,eff}\lesssim \frac{3}{4\pi}\left(\frac{\pi\theta_{jet}^2m}{A_{eff}T_{exp}L_{E*}}\right)^{3/2}
\end{equation}
 imposes an upper bound on a combination $\rho_{*,eff}L_{E*}^{3/2}$ of the source density and luminosity, as discussed in Ref. \cite{waxman}.
 
\begin{figure}
\includegraphics[width=\linewidth]{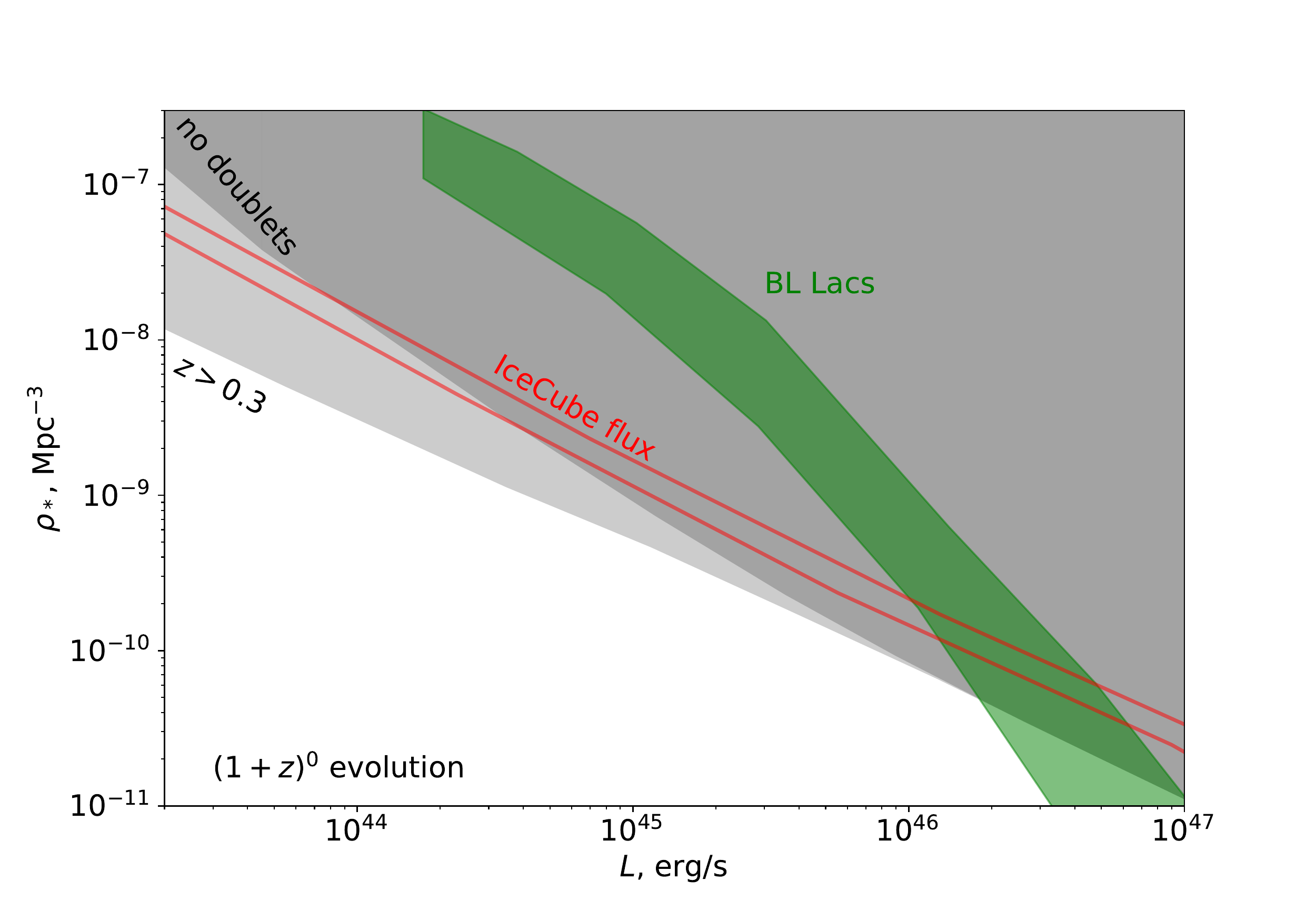}
\caption{95\% confidence level constraints on the properties of non-evolving standard candle neutrino source population. Dark grey shading shows constraint from non-observation of doublets in the muon neutrino sample. Light grey shows constraint from non-observation of neutrino-emitting blazars within redshift $z<0.3$. Red band shows the source density required for production of the observed muon neutrino flux.  }
\label{fig:sc_noevol}
\end{figure}

This analytical result which neglects the Poisson nature of the signal is confirmed by the Monte-Carlo simulation results shown in Fig. \ref{fig:sc_noevol}. The boundary of the dark grey shaded band follows the  $\rho_{*,eff}L_{E*}^{3/2}\sim const$ dependence at large source densities, but deviates from it at low densities, where the sources become sparse and Poisson fluctuations of the signal becomes more important. The $x$ axis of the figure shows the bolometric luminosity $L=\int_{200\ TeV/\epsilon}^\infty L_EdE$.

Absence of identified sources at the distances closer than $z=0.3$ imposes an additional constraint which becomes stronger than the constraint from the absence of doublets at high source densities. Qualitative explanation for this fact is that as the source density grows, it becomes more and more difficult not to notice very nearby sources (which we assume are are all identified as blazars using techniques of multi-wavelength astronomy). Nevertheless, also this constraint shows dependence on the source luminosity because the individual sources get weaker and weak nearby sources on average contribute with less than one neutrino to the signal. The first identifiable source which occasionally produces one event in a given exposure is not necessarily the nearest one. 

A combination of the absence of doublets and absence of nearby sources constraints rules out the possibility that the IceCube muon neutrino flux  is generated by a population of non-evolving  sources, like Low-luminosity BL Lac and Fanaroff-Riley type I  (FR I) radio galaxies which show no or negative cosmological evolution \cite{ajello09_bllac,fri}. This is clear from comparison of the constraints with the density of the sources required for generation of the observed neutrino flux, shown as the red band in Fig. \ref{fig:sc_noevol}. The red band is never found within the allowed range of $\rho_*,L$.

\subsection{Evolving sources}

The constraints from non-observation of doublets and nearby sources are relaxed if the source population is assumed to evolve positively with the redshift, similarly to  high-luminosity BL Lacs and/or FSRQ (and, possibly their progenitors, high luminosity FR II type radio galaxies) \cite{ajello09_bllac,ajello09_fsrq,fri,frii} which  evolve as fast as   $\zeta=5$  up to $z_*\sim 1...2$. 
We consider the possibility of such evolution for neutrino sources  in this section.

The fact that the constraints imposed by non-observation of doublets are weakened for fast evolving sources was already noticed in Ref. \cite{waxman}. The same is true for the constraint from non-observation of nearby sources. This is explained by the fact that in the evolving source population scenario the bulk of the neutrino flux was generated at high redshifts by distant sources. Even though those distant sources were brighter, their individual contributions to the overall neutrino flux observed today are small, with on average much less than one event expected per source. The nearby sources are still weaker than in the non-evolving source population and the probability that nearby source contributes with one neutrino into the signal within a given exposure is still smaller. Only the collective flux from the entire source population is detectable in the form of diffuse emission. 

\begin{figure}
\includegraphics[width=\linewidth]{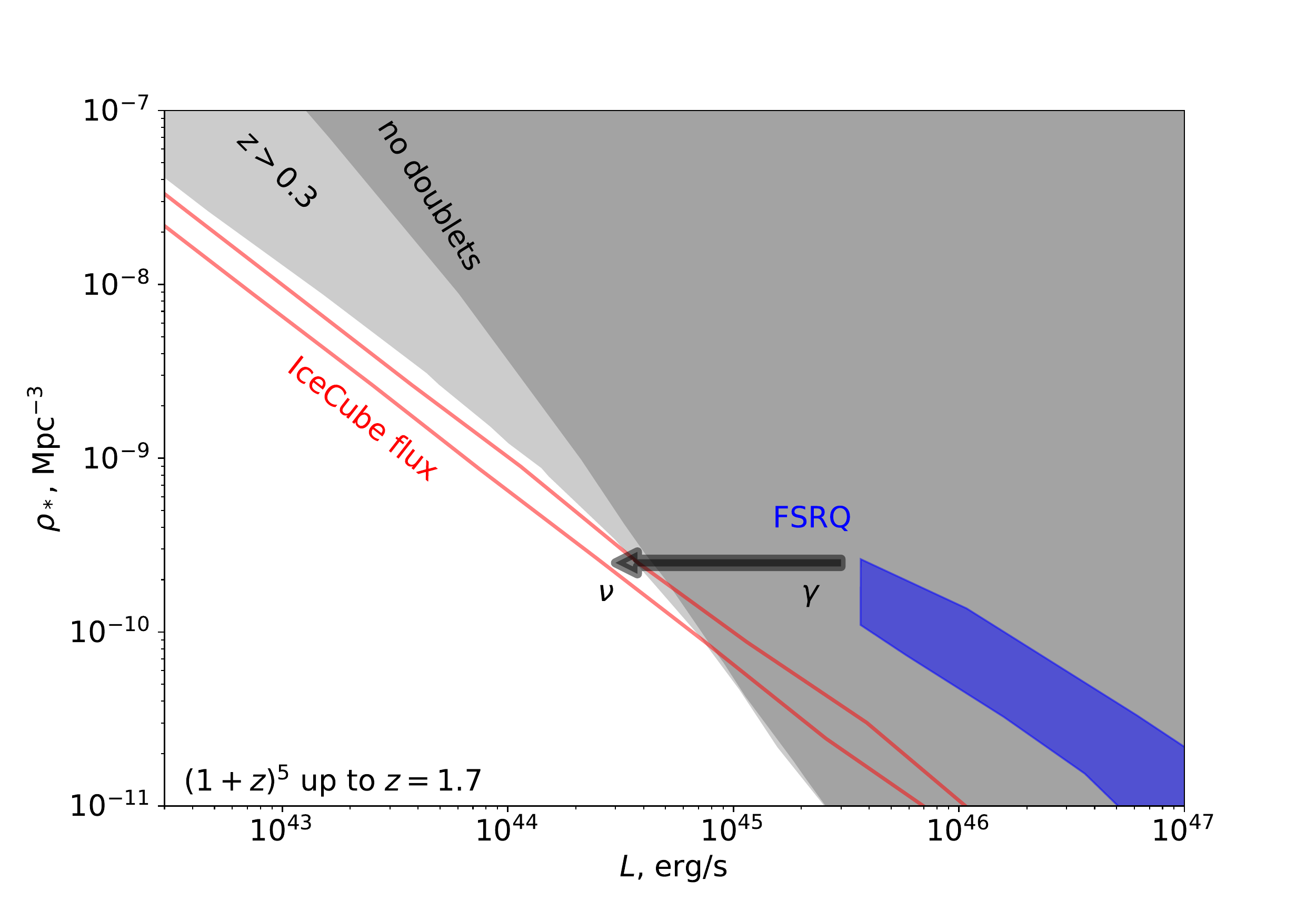}
\caption{Same as in Fig. \ref{fig:sc_noevol} but for  sources evolving with $\zeta=5$ up to redshift $z=1.7$ \cite{hasinger05}.  For comparison, luminosity dependent density of FSRQs (blue shading) from Ref. \cite{ajello09_fsrq} is shown.  }
\label{fig:sc_z3}
\end{figure}

\begin{figure}
\includegraphics[width=\linewidth]{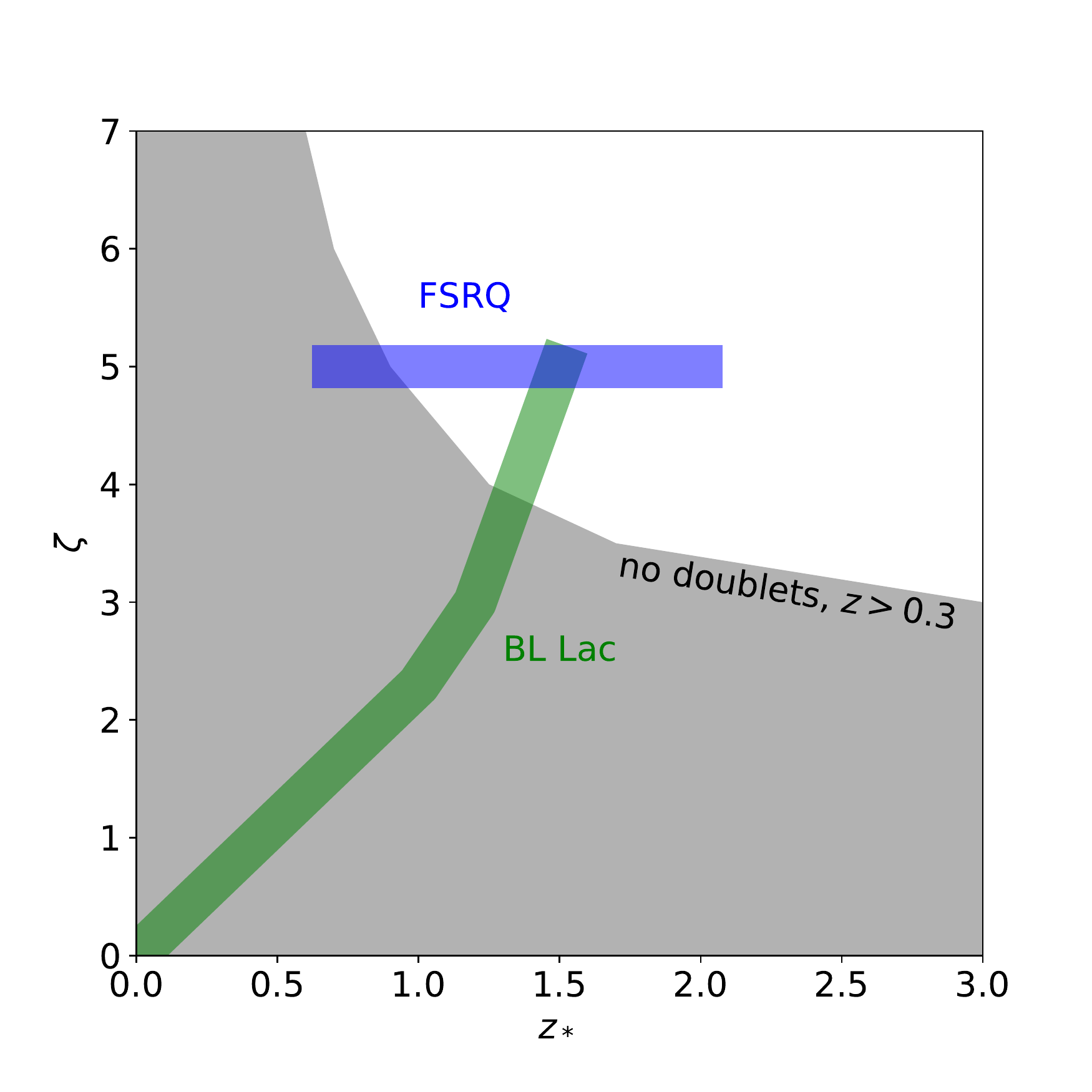}
\caption{Allowed region of evolution parameters for source populations (white area) compared to the evolution parameters of BL Lac \cite{ajello09_bllac} and FSRQ populations \cite{ajello09_fsrq}. }
\label{fig:evolution}
\end{figure}

Fig. \ref{fig:sc_z3} shows that for fast enough evolution with $\zeta\ge 5$ the combined constraint from non-observation of doublets and nearby sources do not rule out a range of source densities needed to provide the observed IceCube muon neutrino flux. Fig. \ref{fig:evolution} shows the allowed range of evolution parameters $z_*$, $\zeta$ within which source population could explain the IceCube muon neutrino flux avoiding constraints from non-identification of nearby sources and absence of doublets in IceCube dataset.   

\section{Constraints on population of neutrino emitting blazars}

Comparison of the properties of the populations of gamma-ray and neutirno emitting blazars is shown in Figs. \ref{fig:sc_noevol}, \ref{fig:sc_z3} and \ref{fig:evolution} where we have overplotted the luminosity function and evolution parameters of BL Lacs from Ref. \cite{ajello09_bllac} and FSRQs from Ref. \cite{ajello09_fsrq}. One could see that sources evolving as majority of BL Lacs could not explain the IceCube signal. Only the evolution parameters of the highest luminosity BL Lacs ($\zeta>4$, $z_*>1.5$) become consistent with IceCube data. However such evolution is valid only for BL Lacs with gamma-ray luminosities in excess of $10^{47}$~erg/s. From Fig. \ref{fig:sc_noevol} one could see that the density of those sources is very low, $n\sim 10^{-11}$~Mpc$^{-3}$.  

To the contrary, Fig. \ref{fig:evolution} shows that the evolution parameters the FSRQ  population are consistent with constraints on evolution parameters of neutrino sources. From Fig. \ref{fig:sc_z3} one could judge that the density of FSRQs is comparable with minimal required density of neutrino sources evolving similarly to FSRQ population. This shows that FSRQs could be considered as viable neutrino source candidates. 

FSRQs are less abundant in the low-redshift Universe than BL Lacs. The closest FSRQ, 3C 273, is at the redshift $z\simeq 0.16$ \cite{3c273}. There are only about 10 FSRQs within the redshift $<0.3$ detected by Fermi/LAT \cite{fermi_catalog} and 3C 273 is brightest among them. The neutrino emitting blazar TXS 0506+056 is classified as BL Lac in SIMBAD astronomical database, but its luminosity scale is closer to that of FSRQs.

\begin{figure}
\includegraphics[width=\linewidth]{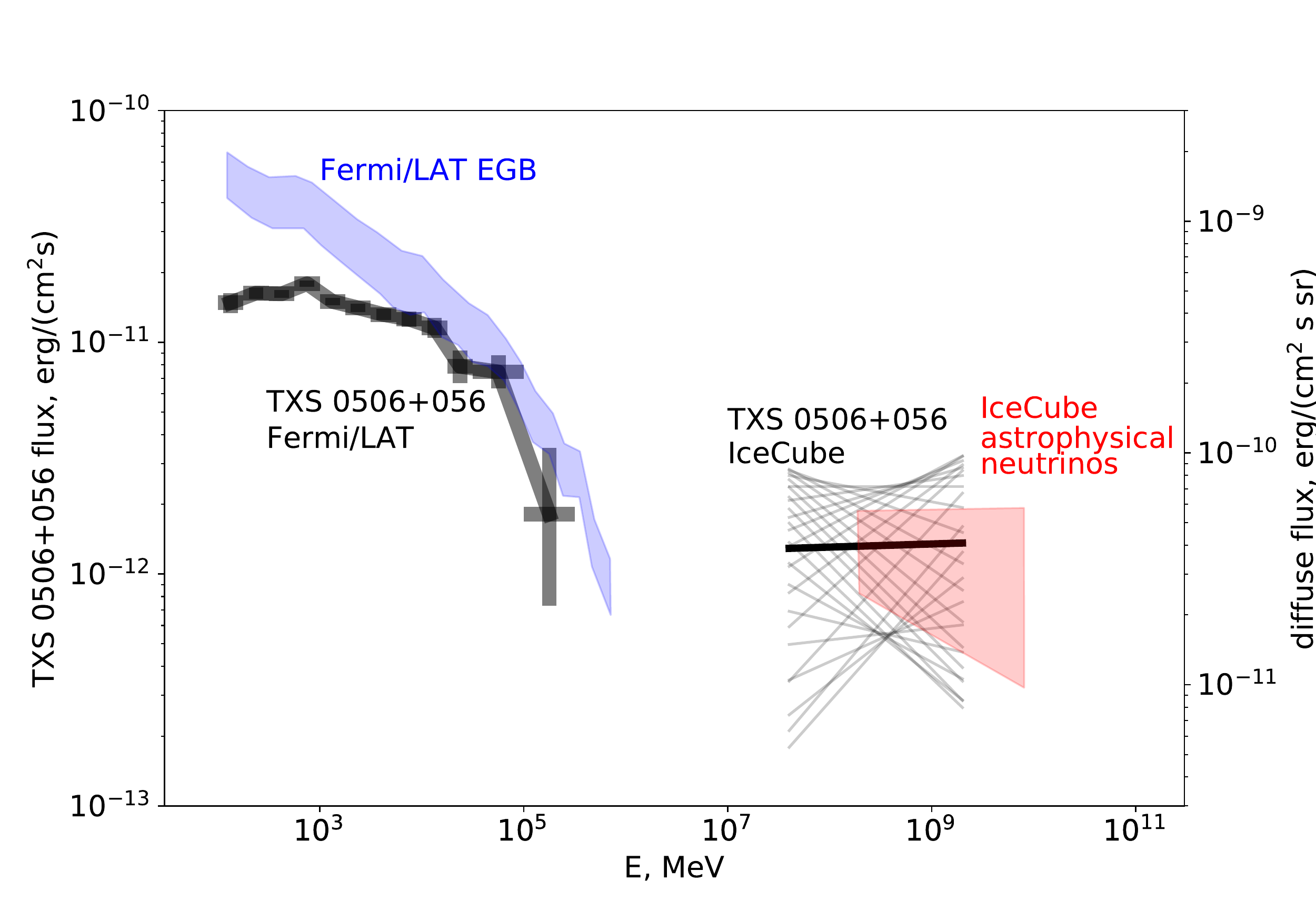}
\caption{Multi-messenger time-averaged spectrum of TXS 0506+056 measured by  IceCube \cite{0506_icecube} (butterfly and black horizontal line) and  Fermi/LAT (grey data points).}
\label{fig:spectrum} 
\end{figure}

Although the evolution parameters and density of FSRQs are consistent with what required for the neutrino sources, their gamma-ray luminosity scale could not be comparable to the neutrino luminosity because in this case they would over-produce the neutrino flux, as it is clear from Fig. \ref{fig:sc_z3}. Neutrino luminosity scale has to be about an order of magnitude below the gamma-ray luminosity. In the simplest model, the neutrino luminosity function consistent with the IceCube measurement could be simply a displacement of the gamma-ray luminosity function along the arrow shown in Fig. \ref{fig:sc_z3}, by an order of magnitude toward lower luminosities.      

TXS 0506+056 multi-messenger detection could provide a useful insight into neutrino-to-gamma-ray luminosity ratio. Its multi-messenger gamma-ray $+$ neutrino spectrum is shown in Fig. \ref{fig:spectrum}. To produce this figure, we have taken the estimate of the time-averaged neutrino flux from the source from Ref. \cite{0506_icecube} and complemented it with the Fermi/LAT time averaged spectrum which we have extracted from the LAT data collected between 2008 and 2018 (fully covering the IceCube exposure). We have used SOURCE class event selection and employed standard likelihood analysis technique described at  using the standard likelihood analysis as described at Fermi Science Support Center website https://fermi.gsfc.nasa.gov/ssc/data/analysis/. 

One could see that the time averaged gamma-ray flux of TXS 0506+056 is more than an order-of-magnitude higher than the time-averaged neutrino flux. This fact is consistent with a possibility that neutrino luminosity of neutrino-loud blazars is at the level of $\lesssim 10\%$ of their gamma-ray luminosity. At the same time, the neutrino-to-gamma-ray flux ratio could vary during flaring activity, as observed TXS 0506+056 in 2015 and 2017 flares \cite{0506_multiwavelength,0506_icecube}.
Such variations are expected because gamma-rays are generated by both leptonic and hadronic processes, while neutrinos are due to hadronic processes only \cite{cerutti18,txs_xray,txs_onemore,taylor18,cerutti18_1}. Occasionally, increased interaction rate of high-energy protons during hadronic flares could boost the neutrino-to-gamma-ray flux ratio. Moreover, the neutrino flux from individual sources could even occasionally dominate ove their gamma-ray luminosity \cite{padovani18}, because the anisotropy patterns of neutrino and gamma-ray emission do not need to be the same. Higher energy neutrino emission not affected by electromagnetic cascade effects could be emitted in a cone with smaller opening angle, while lower energy gamma-ray emission could be emitted with comparable overall luminosity, but in a wider solid anlge, so that its flux in the direction of an observer could be lower than the neutrino flux \cite{neronov_semikoz02}.

\begin{figure}
\includegraphics[width=\linewidth]{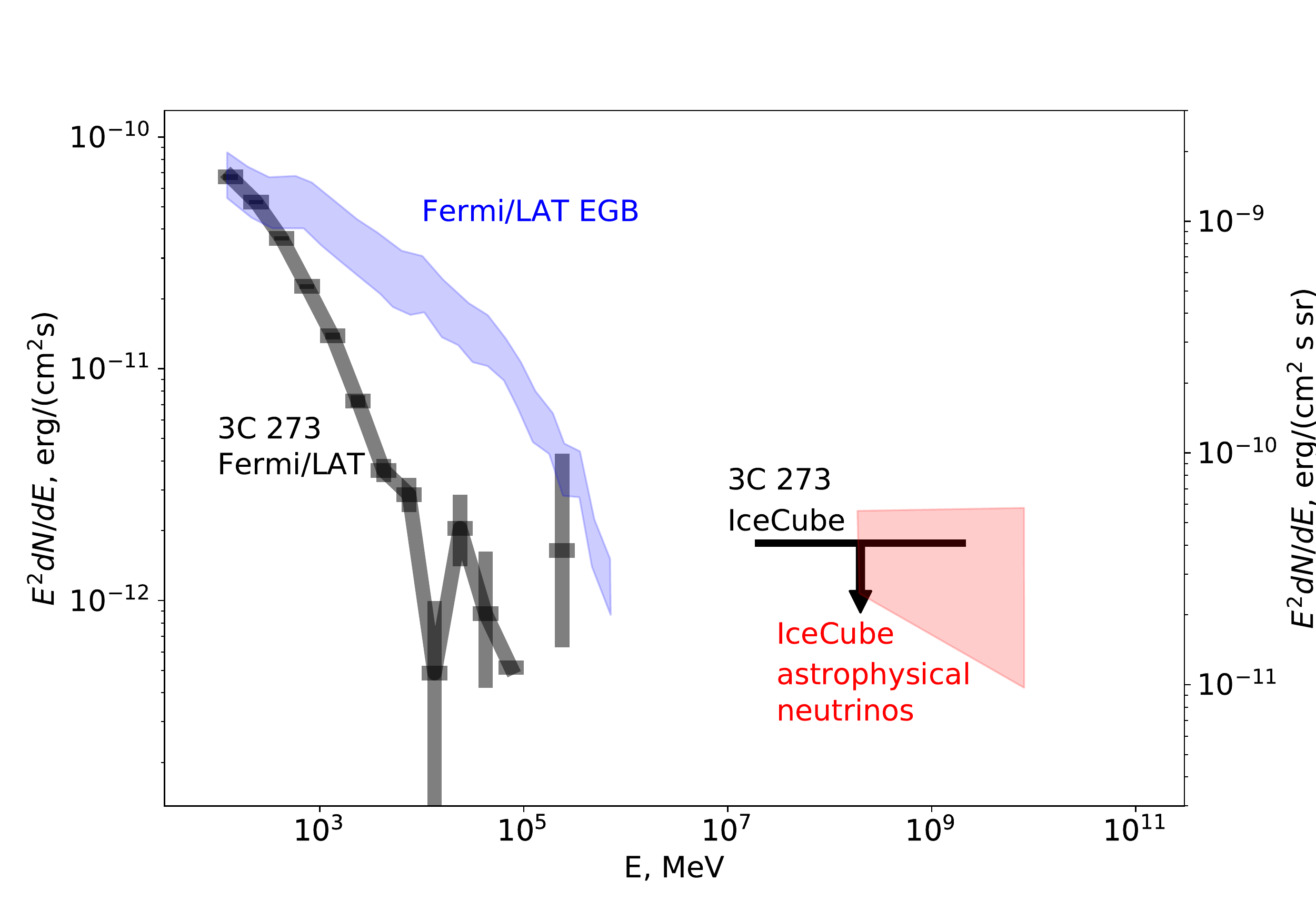}
\caption{Same as in Fig. \ref{fig:spectrum} but for 3C 273.}
\label{fig:spectrum3c273} 
\end{figure}

Fig. \ref{fig:spectrum3c273} shows the multi-messenger spectrum of 3C 273 which includes the upper limit on the neutrino flux derived by IceCube \cite{icecube_pointsources}. From this figure one could see that neutrino flux from 3C 273 is more than an order of magnitude lower than the gamma-ray flux. The upper bound on neutrino flux from 3C 273 is in mild tension with a hypothesis of just an order of magnitude difference between neutrino and gamma-ray luminosities of FSRQs.

Figs. \ref{fig:spectrum}, \ref{fig:spectrum3c273} also show a comparison of the characteristics of the multi-messenger spectra of TXS 0506+056 and 3C 273 with those of the entire neutrino+gamma-ray  extragalactic sky \cite{muon,egbA}. We have chosen the $y$ axis range in such a way that the TXS 0506+056 and 3C 273 gamma-ray fluxes do not exceed the extragalactic gamma-ray flux range. With such y-axis range adjustments, it becomes clear that neutrino-to-gamma-ray flux ratio measurements and upper limit for TXS 0506+056 and 3C 273 are consistent with the neutrino-to-gamma-ray flux ratio of entire extragalactic sky. 

\section{Summary}

Overall, we conclude that the hypothesis of the dominant bright blazar (FSRQ) contribution to the extragalactic neutrino flux is consistent with the observational data (absence of doublets in IceCube muon neutrino sample and $z\sim 0.3$ redshift of the nearest detected source), once the details of cosmological evolution of the source population and differences in the overall luminosity and anisotropy patterns of gamma-ray and neutrino emission are taken into account.

The fact that bright faraway (dominated by FSRQ), rather than dim nearby (dominated by BL Lac) blazars  are neutrino emitters is consistent with a general expectation that neutrino production is more efficient in the presence of dense environment of accretion flow.   FSRQs  host bright radiatively efficient accretion flows which provide abundant target photons for photo-pion production \cite{neronov_semikoz02}. The observational constraints suggest that even though hadronic interactions do not play the dominant role in the energy output of FSRQ, they could still provide a sizeable contribution to the astrophysical neutrino flux if all FSRQs typically produce neutrino flux is at the level up to 10\% of the gamma-ray flux. 

\bibliography{Narrow_jets}

\providecommand{\href}[2]{#2}\begingroup\raggedright\begin{thebibliography}{10}

\bibitem{icecube_discovery}
I.~Collaboration, {\it Evidence for high-energy extraterrestrial neutrinos at
  the icecube detector},  {\em Science} {\bf 342} (2013), no.~6161
  [\href{http://arXiv.org/abs/http://science.sciencemag.org/content/342/6161/1242856.full.pdf}{{\tt
  http://science.sciencemag.org/content/342/6161/1242856.full.pdf}}].

\bibitem{hese}
M.~G. {Aartsen}, M.~{Ackermann}, J.~{Adams}, J.~A. {Aguilar}, M.~{Ahlers},
  M.~{Ahrens}, D.~{Altmann}, T.~{Anderson}, C.~{Arguelles}, T.~C. {Arlen} and
  et~al., {\it {Observation of High-Energy Astrophysical Neutrinos in Three
  Years of IceCube Data}},  {\em Physical Review Letters} {\bf 113} (Sept.,
  2014) 101101 [\href{http://arXiv.org/abs/1405.5303}{{\tt 1405.5303}}].

\bibitem{muon}
M.~G. Aartsen, K.~Abraham, M.~Ackermann, J.~Adams, J.~A. Aguilar, M.~Ahlers,
  M.~Ahrens, D.~Altmann, K.~Andeen, T.~Anderson, I.~Ansseau, G.~Anton,
  M.~Archinger, C.~Argüelles, J.~Auffenberg, S.~Axani, X.~Bai, S.~W. Barwick,
  V.~Baum, R.~Bay, J.~J. Beatty, J.~B. Tjus, K.-H. Becker, S.~BenZvi,
  P.~Berghaus, D.~Berley, E. and I.~Collaboration), {\it Observation and
  characterization of a cosmic muon neutrino flux from the northern hemisphere
  using six years of icecube data},  {\em The Astrophysical Journal} {\bf 833}
  (2016), no.~1 3.

\bibitem{neronov_semikoz}
A.~{Neronov}, M.~{Kachelrie{\ss}} and D.~V. {Semikoz}, {\it {Multimessenger
  gamma-ray counterpart of the IceCube neutrino signal}},  {\em \prd} {\bf 98}
  (July, 2018) 023004 [\href{http://arXiv.org/abs/1802.09983}{{\tt
  1802.09983}}].

\bibitem{neronov_semikoz_muon}
A.~{Neronov} and D.~{Semikoz}, {\it {Galactic and extragalactic contributions
  to the astrophysical muon neutrino signal}},  {\em \prd} {\bf 93} (June,
  2016) 123002 [\href{http://arXiv.org/abs/1603.06733}{{\tt 1603.06733}}].

\bibitem{neronov_semikoz_evidence}
A.~{Neronov} and D.~{Semikoz}, {\it {Evidence the Galactic contribution to the
  IceCube astrophysical neutrino flux}},  {\em Astroparticle Physics} {\bf 75}
  (Feb., 2016) 60--63 [\href{http://arXiv.org/abs/1509.03522}{{\tt
  1509.03522}}].

\bibitem{neronov_semikoz_lhc}
A.~{Neronov} and D.~{Semikoz}, {\it {Neutrinos from extra-large Hadron Collider
  in the Milky Way}},  {\em Astroparticle Physics} {\bf 72} (Jan., 2016) 32--37
  [\href{http://arXiv.org/abs/1412.1690}{{\tt 1412.1690}}].

\bibitem{neronov_semikoz_galactic1}
A.~{Neronov}, D.~{Semikoz} and C.~{Tchernin}, {\it {PeV neutrinos from
  interactions of cosmic rays with the interstellar medium in the Galaxy}},
  {\em \prd} {\bf 89} (May, 2014) 103002
  [\href{http://arXiv.org/abs/1307.2158}{{\tt 1307.2158}}].

\bibitem{galactic_icecube}
M.~G. {Aartsen}, M.~{Ackermann}, J.~{Adams}, J.~A. {Aguilar}, M.~{Ahlers},
  M.~{Ahrens}, I.~A. {Samarai}, D.~{Altmann}, K.~{Andeen}, T.~{Anderson} and
  et~al., {\it {Constraints on Galactic Neutrino Emission with Seven Years of
  IceCube Data}},  {\em \apj} {\bf 849} (Nov., 2017) 67
  [\href{http://arXiv.org/abs/1707.03416}{{\tt 1707.03416}}].

\bibitem{galactic_icecube_antares}
A.~{Albert}, M.~{Andr{\'e}}, M.~{Anghinolfi}, M.~{Ardid}, J.-J. {Aubert},
  J.~{Aublin}, T.~{Avgitas}, B.~{Baret}, J.~{Barrios-Mart{\'{\i}}}, S.~{Basa}
  and et~al., {\it {Joint constraints on Galactic diffuse neutrino emission
  from ANTARES and IceCube}},  {\em ArXiv e-prints} (Aug., 2018)
  [\href{http://arXiv.org/abs/1808.03531}{{\tt 1808.03531}}].

\bibitem{galactic_antares}
A.~{Albert}, M.~{Andr{\'e}}, M.~{Anghinolfi}, G.~{Anton}, M.~{Ardid}, J.-J.
  {Aubert}, T.~{Avgitas}, B.~{Baret}, J.~{Barrios-Mart{\'{\i}}}, S.~{Basa},
  B.~{Belhorma}, V.~{Bertin}, S.~{Biagi}, R.~{Bormuth}, S.~{Bourret}, M.~C.
  {Bouwhuis}, R.~{Bruijn}, J.~{Brunner}, J.~{Busto}, A.~{Capone},
  L.~{Caramete}, J.~{Carr}, S.~{Celli}, R.~{Cherkaoui El Moursli},
  T.~{Chiarusi}, M.~{Circella}, J.~A.~B. {Coelho}, A.~{Coleiro},
  R.~{Coniglione}, H.~{Costantini}, P.~{Coyle}, A.~{Creusot}, A.~F.
  {D{\'{\i}}az}, A.~{Deschamps}, G.~{de Bonis}, C.~{Distefano}, I.~{di Palma},
  A.~{Domi}, C.~{Donzaud}, D.~{Dornic}, D.~{Drouhin}, T.~{Eberl}, I.~{El
  Bojaddaini}, N.~{El Khayati}, D.~{Els{\"a}sser}, A.~{Enzenh{\"o}fer},
  A.~{Ettahiri}, F.~{Fassi}, I.~{Felis}, L.~A. {Fusco}, S.~{Galat{\`a}},
  P.~{Gay}, V.~{Giordano}, H.~{Glotin}, T.~{Gr{\'e}goire}, R.~{Gracia Ruiz},
  K.~{Graf}, S.~{Hallmann}, H.~{van Haren}, A.~J. {Heijboer}, Y.~{Hello}, J.~J.
  {Hern{\'a}ndez-Rey}, J.~{H{\"o}{\ss}l}, J.~{Hofest{\"a}dt}, C.~{Hugon},
  G.~{Illuminati}, C.~W. {James}, M.~{de Jong}, M.~{Jongen}, M.~{Kadler},
  O.~{Kalekin}, U.~{Katz}, D.~{Kie{\ss}ling}, A.~{Kouchner}, M.~{Kreter},
  I.~{Kreykenbohm}, V.~{Kulikovskiy}, C.~{Lachaud}, R.~{Lahmann},
  D.~{Lef{\`e}vre}, E.~{Leonora}, M.~{Lotze}, S.~{Loucatos}, M.~{Marcelin},
  A.~{Margiotta}, A.~{Marinelli}, J.~A. {Mart{\'{\i}}nez-Mora}, R.~{Mele},
  K.~{Melis}, T.~{Michael}, P.~{Migliozzi}, A.~{Moussa}, S.~{Navas},
  E.~{Nezri}, M.~{Organokov}, G.~E. {P{\v a}v{\v a}la{\c s}}, C.~{Pellegrino},
  C.~{Perrina}, P.~{Piattelli}, V.~{Popa}, T.~{Pradier}, L.~{Quinn},
  C.~{Racca}, G.~{Riccobene}, A.~{S{\'a}nchez-Losa}, M.~{Salda{\~n}a},
  I.~{Salvadori}, D.~F.~E. {Samtleben}, M.~{Sanguineti}, P.~{Sapienza},
  F.~{Sch{\"u}ssler}, C.~{Sieger}, M.~{Spurio}, T.~{Stolarczyk}, M.~{Taiuti},
  Y.~{Tayalati}, A.~{Trovato}, D.~{Turpin}, C.~{T{\"o}nnis}, B.~{Vallage},
  V.~{van Elewyck}, F.~{Versari}, D.~{Vivolo}, A.~{Vizzoca}, J.~{Wilms}, J.~D.
  {Zornoza}, J.~{Z{\'u}{\~n}iga}, D.~{Gaggero}, D.~{Grasso} and {ANTARES
  Collaboration}, {\it {New constraints on all flavor Galactic diffuse neutrino
  emission with the ANTARES telescope}},  {\em \prd} {\bf 96} (Sept., 2017)
  062001 [\href{http://arXiv.org/abs/1705.00497}{{\tt 1705.00497}}].

\bibitem{giacinti18}
G.~{Giacinti}, M.~{Kachelrie{\ss}} and D.~V. {Semikoz}, {\it {Reconciling
  cosmic ray diffusion with Galactic magnetic field models}},  {\em \jcap} {\bf
  7} (July, 2018) 051 [\href{http://arXiv.org/abs/1710.08205}{{\tt
  1710.08205}}].

\bibitem{icecube_combined}
M.~G. Aartsen, K.~Abraham, M.~Ackermann, J.~Adams, J.~A. Aguilar, M.~Ahlers,
  M.~Ahrens, D.~Altmann, T.~Anderson, M.~Archinger, C.~Arguelles, T.~C. Arlen,
  J.~Auffenberg, X.~Bai, S.~W. Barwick, V.~Baum, R.~Bay and T.~I.
  Collaboration, {\it A combined maximum-likelihood analysis of the high-energy
  astrophysical neutrino flux measured with icecube},  {\em The Astrophysical
  Journal} {\bf 809} (2015), no.~1 98.

\bibitem{wb}
E.~{Waxman} and J.~{Bahcall}, {\it {High energy neutrinos from astrophysical
  sources: An upper bound}},  {\em \prd} {\bf 59} (Jan., 1999) 023002
  [\href{http://arXiv.org/abs/hep-ph/9807282}{{\tt hep-ph/9807282}}].

\bibitem{giacinti15}
G.~{Giacinti}, M.~{Kachelrie{\ss}}, O.~{Kalashev}, A.~{Neronov} and D.~V.
  {Semikoz}, {\it {Unified model for cosmic rays above 10$^{17}$ eV and the
  diffuse gamma-ray and neutrino backgrounds}},  {\em \prd} {\bf 92} (Oct.,
  2015) 083016 [\href{http://arXiv.org/abs/1507.07534}{{\tt 1507.07534}}].

\bibitem{kalashev17}
M.~{Kachelrie{\ss}}, O.~{Kalashev}, S.~{Ostapchenko} and D.~V. {Semikoz}, {\it
  {Minimal model for extragalactic cosmic rays and neutrinos}},  {\em \prd}
  {\bf 96} (Oct., 2017) 083006 [\href{http://arXiv.org/abs/1704.06893}{{\tt
  1704.06893}}].

\bibitem{berezinsky69}
V.~S. {Beresinsky} and G.~T. {Zatsepin}, {\it {Cosmic rays at ultra high
  energies (neutrino?)}},  {\em Physics Letters B} {\bf 28} (Jan., 1969)
  423--424.

\bibitem{berezinsky75}
V.~S. {Berezinskii} and A.~I. {Smirnov}, {\it {Cosmic neutrinos of ultra-high
  energies and detection possibility}},  {\em \apss} {\bf 32} (Feb., 1975)
  461--482.

\bibitem{neronov_semikoz02}
A.~Y. {Neronov} and D.~V. {Semikoz}, {\it {Which blazars are neutrino loud?}},
  {\em \prd} {\bf 66} (Dec., 2002) 123003
  [\href{http://arXiv.org/abs/hep-ph/0208248}{{\tt hep-ph/0208248}}].

\bibitem{neronov_sibiryakov}
A.~{Neronov}, D.~{Semikoz} and S.~{Sibiryakov}, {\it {Measuring parameters of
  active galactic nuclei central engines with very high energy {$\gamma$}-ray
  flares}},  {\em \mnras} {\bf 391} (Dec., 2008) 949--958
  [\href{http://arXiv.org/abs/0806.2545}{{\tt 0806.2545}}].

\bibitem{njp}
A.~Y. {Neronov}, D.~V. {Semikoz} and I.~I. {Tkachev}, {\it {Ultra-high energy
  cosmic ray production in the polar cap regions of black hole
  magnetospheres}},  {\em New Journal of Physics} {\bf 11} (June, 2009) 065015
  [\href{http://arXiv.org/abs/0712.1737}{{\tt 0712.1737}}].

\bibitem{mb89}
K.~{Mannheim} and P.~L. {Biermann}, {\it {Photomeson production in active
  galactic nuclei}},  {\em \aap} {\bf 221} (Sept., 1989) 211--220.

\bibitem{mannheim}
K.~{Mannheim}, {\it {The proton blazar}},  {\em \aap} {\bf 269} (Mar., 1993)
  67--76 [\href{http://arXiv.org/abs/astro-ph/9302006}{{\tt
  astro-ph/9302006}}].

\bibitem{tchernin}
C.~{Tchernin}, J.~A. {Aguilar}, A.~{Neronov} and T.~{Montaruli}, {\it {An
  exploration of hadronic interactions in blazars using IceCube}},  {\em \aap}
  {\bf 555} (July, 2013) A70 [\href{http://arXiv.org/abs/1305.3524}{{\tt
  1305.3524}}].

\bibitem{cerutti18}
M.~{Cerruti}, A.~{Zech}, C.~{Boisson}, G.~{Emery}, S.~{Inoue} and J.-P.
  {Lenain}, {\it {Gammas and neutrinos from TXS 0506+056}},  {\em ArXiv
  e-prints} (Oct., 2018) [\href{http://arXiv.org/abs/1810.08825}{{\tt
  1810.08825}}].

\bibitem{cerutti18_1}
M.~{Cerruti}, A.~{Zech}, C.~{Boisson}, G.~{Emery}, S.~{Inoue} and J.-P.
  {Lenain}, {\it {Lepto-hadronic single-zone models for the electromagnetic and
  neutrino emission of TXS 0506+056}},  {\em ArXiv e-prints} (July, 2018)
  [\href{http://arXiv.org/abs/1807.04335}{{\tt 1807.04335}}].

\bibitem{taylor18}
R.-Y. {Liu}, K.~{Wang}, R.~{Xue}, A.~M. {Taylor}, X.-Y. {Wang}, Z.~{Li} and
  H.~{Yan}, {\it {A hadronuclear interpretation of a high-energy neutrino event
  coincident with a blazar flare}},  {\em ArXiv e-prints} (July, 2018)
  [\href{http://arXiv.org/abs/1807.05113}{{\tt 1807.05113}}].

\bibitem{tavecchio18}
C.~{Righi}, F.~{Tavecchio} and S.~{Inoue}, {\it {Neutrino emission from BL Lac
  objects: the role of radiatively inefficient accretion flows}},  {\em ArXiv
  e-prints} (July, 2018) [\href{http://arXiv.org/abs/1807.10506}{{\tt
  1807.10506}}].

\bibitem{padovani18}
P.~{Padovani}, P.~{Giommi}, E.~{Resconi}, T.~{Glauch}, B.~{Arsioli},
  N.~{Sahakyan} and M.~{Huber}, {\it {Dissecting the region around
  IceCube-170922A: the blazar TXS 0506+056 as the first cosmic neutrino
  source}},  {\em \mnras} {\bf 480} (Oct., 2018) 192--203
  [\href{http://arXiv.org/abs/1807.04461}{{\tt 1807.04461}}].

\bibitem{txs_xray}
A.~{Keivani}, K.~{Murase}, M.~{Petropoulou}, D.~B. {Fox}, S.~B. {Cenko},
  S.~{Chaty}, A.~{Coleiro}, J.~J. {DeLaunay}, S.~{Dimitrakoudis}, P.~A.
  {Evans}, J.~A. {Kennea}, F.~E. {Marshall}, A.~{Mastichiadis}, J.~P.
  {Osborne}, M.~{Santander}, A.~{Tohuvavohu} and C.~F. {Turley}, {\it {A
  Multimessenger Picture of the Flaring Blazar TXS 0506+056: Implications for
  High-energy Neutrino Emission and Cosmic-Ray Acceleration}},  {\em \apj} {\bf
  864} (Sept., 2018) 84 [\href{http://arXiv.org/abs/1807.04537}{{\tt
  1807.04537}}].

\bibitem{txs_onemore}
S.~{Ansoldi}, L.~A. {Antonelli}, C.~{Arcaro}, D.~{Baack}, A.~{Babi{\'c}},
  B.~{Banerjee}, P.~{Bangale}, U.~{Barres de Almeida}, J.~A. {Barrio},
  J.~{Becerra Gonz{\'a}lez}, W.~{Bednarek}, E.~{Bernardini}, R.~C. {Berse},
  A.~{Berti}, J.~{Besenrieder}, W.~{Bhattacharyya}, C.~{Bigongiari},
  A.~{Biland}, O.~{Blanch}, G.~{Bonnoli}, R.~{Carosi}, G.~{Ceribella},
  A.~{Chatterjee}, S.~M. {Colak}, P.~{Colin}, E.~{Colombo}, J.~L. {Contreras},
  J.~{Cortina}, S.~{Covino}, P.~{Cumani}, V.~{D'Elia}, P.~{Da Vela},
  F.~{Dazzi}, A.~{De Angelis}, B.~{De Lotto}, M.~{Delfino}, J.~{Delgado},
  F.~{Di Pierro}, A.~{Dom{\'{\i}}nguez}, D.~{Dominis Prester}, D.~{Dorner},
  M.~{Doro}, S.~{Einecke}, D.~{Elsaesser}, V.~{Fallah Ramazani},
  A.~{Fattorini}, A.~{Fern{\'a}ndez-Barral}, G.~{Ferrara}, D.~{Fidalgo},
  L.~{Foffano}, M.~V. {Fonseca}, L.~{Font}, C.~{Fruck}, S.~{Gallozzi}, R.~J.
  {Garc{\'{\i}}a L{\'o}pez}, M.~{Garczarczyk}, M.~{Gaug}, P.~{Giammaria},
  N.~{Godinovi{\'c}}, D.~{Guberman}, D.~{Hadasch}, A.~{Hahn}, T.~{Hassan},
  M.~{Hayashida}, J.~{Herrera}, J.~{Hoang}, D.~{Hrupec}, S.~{Inoue},
  K.~{Ishio}, Y.~{Iwamura}, Y.~{Konno}, H.~{Kubo}, J.~{Kushida}, A.~{Lamastra},
  D.~{Lelas}, F.~{Leone}, E.~{Lindfors}, S.~{Lombardi}, F.~{Longo},
  M.~{L{\'o}pez}, C.~{Maggio}, P.~{Majumdar}, M.~{Makariev}, G.~{Maneva},
  M.~{Manganaro}, K.~{Mannheim}, L.~{Maraschi}, M.~{Mariotti},
  M.~{Mart{\'{\i}}nez}, S.~{Masuda}, D.~{Mazin}, K.~{Mielke}, M.~{Minev}, J.~M.
  {Miranda}, R.~{Mirzoyan}, A.~{Moralejo}, V.~{Moreno}, E.~{Moretti},
  V.~{Neustroev}, A.~{Niedzwiecki}, M.~{Nievas Rosillo}, C.~{Nigro},
  K.~{Nilsson}, D.~{Ninci}, K.~{Nishijima}, K.~{Noda}, L.~{Nogu{\'e}s},
  S.~{Paiano}, J.~{Palacio}, D.~{Paneque}, R.~{Paoletti}, J.~M. {Paredes},
  G.~{Pedaletti}, P.~{Pe{\~n}il}, M.~{Peresano}, M.~{Persic}, K.~{Pfrang},
  P.~G. {Prada Moroni}, E.~{Prandini}, I.~{Puljak}, J.~R. {Garcia}, W.~{Rhode},
  M.~{Rib{\'o}}, J.~{Rico}, C.~{Righi}, A.~{Rugliancich}, L.~{Saha},
  T.~{Saito}, K.~{Satalecka}, T.~{Schweizer}, J.~{Sitarek}, I.~{{\v
  S}nidari{\'c}}, D.~{Sobczynska}, A.~{Stamerra}, M.~{Strzys}, T.~{Suri{\'c}},
  F.~{Tavecchio}, P.~{Temnikov}, T.~{Terzi{\'c}}, M.~{Teshima},
  N.~{Torres-Alb{\'a}}, S.~{Tsujimoto}, G.~{Vanzo}, M.~{Vazquez Acosta},
  I.~{Vovk}, J.~E. {Ward}, M.~{Will}, D.~{Zari{\'c}} and M.~{Cerruti}, {\it
  {The Blazar TXS 0506+056 Associated with a High-energy Neutrino: Insights
  into Extragalactic Jets and Cosmic-Ray Acceleration}},  {\em \apjl} {\bf 863}
  (Aug., 2018) L10 [\href{http://arXiv.org/abs/1807.04300}{{\tt 1807.04300}}].

\bibitem{neronov_semikoz_blazars}
A.~{Neronov}, D.~V. {Semikoz} and K.~{Ptitsyna}, {\it {Strong constraints on
  hadronic models of blazar activity from Fermi and IceCube stacking
  analysis}},  {\em \aap} {\bf 603} (July, 2017) A135
  [\href{http://arXiv.org/abs/1611.06338}{{\tt 1611.06338}}].

\bibitem{icecube_blazars}
M.~G. {Aartsen}, K.~{Abraham}, M.~{Ackermann}, J.~{Adams}, J.~A. {Aguilar},
  M.~{Ahlers}, M.~{Ahrens}, D.~{Altmann}, K.~{Andeen}, T.~{Anderson} and
  et~al., {\it {The Contribution of Fermi-2LAC Blazars to Diffuse TeV-PeV
  Neutrino Flux}},  {\em \apj} {\bf 835} (Jan., 2017) 45
  [\href{http://arXiv.org/abs/1611.03874}{{\tt 1611.03874}}].

\bibitem{0506_multiwavelength}
M.~Aartsen, M.~Ackermann, J.~Adams, J.~A. Aguilar, M.~Ahlers, M.~Ahrens,
  I.~Al~Samarai, D.~Altmann, K.~Andeen, T.~Anderson, I.~Ansseau, G.~Anton,
  C.~Arg{\"u}elles, J.~Auffenberg, S.~Axani and e.~a. Bagherpour, {\it
  Multimessenger observations of a flaring blazar coincident with high-energy
  neutrino icecube-170922a},  {\em Science} {\bf 361} (2018), no.~6398
  [\href{http://arXiv.org/abs/http://science.sciencemag.org/content/361/6398/eaat1378.full.pdf}{{\tt
  http://science.sciencemag.org/content/361/6398/eaat1378.full.pdf}}].

\bibitem{0506_icecube}
M.~Aartsen, M.~Ackermann, J.~Adams, J.~A. Aguilar, M.~Ahlers, M.~Ahrens and
  A.~S. et~al., {\it Neutrino emission from the direction of the blazar txs
  0506+056 prior to the icecube-170922a alert},  {\em Science} (2018)
  [\href{http://arXiv.org/abs/http://science.sciencemag.org/content/early/2018/07/11/science.aat2890.full.pdf}{{\tt
  http://science.sciencemag.org/content/early/2018/07/11/science.aat2890.full.pdf}}].

\bibitem{waxman}
K.~{Murase} and E.~{Waxman}, {\it {Constraining high-energy cosmic neutrino
  sources: Implications and prospects}},  {\em \prd} {\bf 94} (Nov., 2016)
  103006 [\href{http://arXiv.org/abs/1607.01601}{{\tt 1607.01601}}].

\bibitem{0506_redshift}
S.~{Paiano}, R.~{Falomo}, A.~{Treves} and R.~{Scarpa}, {\it {The Redshift of
  the BL Lac Object TXS 0506+056}},  {\em \apjl} {\bf 854} (Feb., 2018) L32
  [\href{http://arXiv.org/abs/1802.01939}{{\tt 1802.01939}}].

\bibitem{ajello09_fsrq}
M.~{Ajello}, M.~S. {Shaw}, R.~W. {Romani}, C.~D. {Dermer}, L.~{Costamante},
  O.~G. {King}, W.~{Max-Moerbeck}, A.~{Readhead}, A.~{Reimer}, J.~L. {Richards}
  and M.~{Stevenson}, {\it {The Luminosity Function of Fermi-detected
  Flat-spectrum Radio Quasars}},  {\em \apj} {\bf 751} (June, 2012) 108
  [\href{http://arXiv.org/abs/1110.3787}{{\tt 1110.3787}}].

\bibitem{ajello09_bllac}
M.~{Ajello}, R.~W. {Romani}, D.~{Gasparrini}, M.~S. {Shaw}, J.~{Bolmer},
  G.~{Cotter}, J.~{Finke}, J.~{Greiner}, S.~E. {Healey}, O.~{King},
  W.~{Max-Moerbeck}, P.~F. {Michelson}, W.~J. {Potter}, A.~{Rau}, A.~C.~S.
  {Readhead}, J.~L. {Richards} and P.~{Schady}, {\it {The Cosmic Evolution of
  Fermi BL Lacertae Objects}},  {\em \apj} {\bf 780} (Jan., 2014) 73
  [\href{http://arXiv.org/abs/1310.0006}{{\tt 1310.0006}}].

\bibitem{fri}
E.~M. {Sadler}, R.~D. {Cannon}, T.~{Mauch}, P.~J. {Hancock}, D.~A. {Wake},
  N.~{Ross}, S.~M. {Croom}, M.~J. {Drinkwater}, A.~C. {Edge}, D.~{Eisenstein},
  A.~M. {Hopkins}, H.~M. {Johnston}, R.~{Nichol}, K.~A. {Pimbblet}, R.~{de
  Propris}, I.~G. {Roseboom}, D.~P. {Schneider} and T.~{Shanks}, {\it {Radio
  galaxies in the 2SLAQ Luminous Red Galaxy Survey - I. The evolution of
  low-power radio galaxies to z \~{} 0.7}},  {\em \mnras} {\bf 381} (Oct.,
  2007) 211--227 [\href{http://arXiv.org/abs/astro-ph/0612019}{{\tt
  astro-ph/0612019}}].

\bibitem{frii}
V.~{Smol{\v c}i{\'c}}, G.~{Zamorani}, E.~{Schinnerer}, S.~{Bardelli},
  M.~{Bondi}, L.~{B{\^i}rzan}, C.~L. {Carilli}, P.~{Ciliegi}, M.~{Elvis}, C.~D.
  {Impey}, A.~M. {Koekemoer}, A.~{Merloni}, T.~{Paglione}, M.~{Salvato},
  M.~{Scodeggio}, N.~{Scoville} and J.~R. {Trump}, {\it {Cosmic Evolution of
  Radio Selected Active Galactic Nuclei in the Cosmos Field}},  {\em \apj} {\bf
  696} (May, 2009) 24--39 [\href{http://arXiv.org/abs/0901.3372}{{\tt
  0901.3372}}].

\bibitem{hasinger05}
G.~{Hasinger}, T.~{Miyaji} and M.~{Schmidt}, {\it {Luminosity-dependent
  evolution of soft X-ray selected AGN. New Chandra and XMM-Newton surveys}},
  {\em \aap} {\bf 441} (Oct., 2005) 417--434
  [\href{http://arXiv.org/abs/astro-ph/0506118}{{\tt astro-ph/0506118}}].

\bibitem{3c273}
T.~J.-L. {Courvoisier}, {\it {The bright quasar 3C 273}},  {\em \aapr} {\bf 9}
  (1998) 1--32 [\href{http://arXiv.org/abs/astro-ph/9808147}{{\tt
  astro-ph/9808147}}].

\bibitem{fermi_catalog}
F.~{Acero}, M.~{Ackermann}, M.~{Ajello}, A.~{Albert}, W.~B. {Atwood},
  M.~{Axelsson}, L.~{Baldini}, J.~{Ballet}, G.~{Barbiellini}, D.~{Bastieri},
  A.~{Belfiore}, R.~{Bellazzini}, E.~{Bissaldi}, R.~D. {Blandford}, E.~D.
  {Bloom}, J.~R. {Bogart}, R.~{Bonino}, E.~{Bottacini}, J.~{Bregeon}, R.~J.
  {Britto}, P.~{Bruel}, R.~{Buehler}, T.~H. {Burnett}, S.~{Buson}, G.~A.
  {Caliandro}, R.~A. {Cameron}, R.~{Caputo}, M.~{Caragiulo}, P.~A. {Caraveo},
  J.~M. {Casandjian}, E.~{Cavazzuti}, E.~{Charles}, R.~C.~G. {Chaves},
  A.~{Chekhtman}, C.~C. {Cheung}, J.~{Chiang}, G.~{Chiaro}, S.~{Ciprini},
  R.~{Claus}, J.~{Cohen-Tanugi}, L.~R. {Cominsky}, J.~{Conrad}, S.~{Cutini},
  F.~{D'Ammando}, A.~{de Angelis}, M.~{DeKlotz}, F.~{de Palma}, R.~{Desiante},
  S.~W. {Digel}, L.~{Di Venere}, P.~S. {Drell}, R.~{Dubois}, D.~{Dumora},
  C.~{Favuzzi}, S.~J. {Fegan}, E.~C. {Ferrara}, J.~{Finke}, A.~{Franckowiak},
  Y.~{Fukazawa}, S.~{Funk}, P.~{Fusco}, F.~{Gargano}, D.~{Gasparrini},
  B.~{Giebels}, N.~{Giglietto}, P.~{Giommi}, F.~{Giordano}, M.~{Giroletti},
  T.~{Glanzman}, G.~{Godfrey}, I.~A. {Grenier}, M.-H. {Grondin}, J.~E. {Grove},
  L.~{Guillemot}, S.~{Guiriec}, D.~{Hadasch}, A.~K. {Harding}, E.~{Hays}, J.~W.
  {Hewitt}, A.~B. {Hill}, D.~{Horan}, G.~{Iafrate}, T.~{Jogler},
  G.~{J{\'o}hannesson}, R.~P. {Johnson}, A.~S. {Johnson}, T.~J. {Johnson},
  W.~N. {Johnson}, T.~{Kamae}, J.~{Kataoka}, J.~{Katsuta}, M.~{Kuss}, G.~{La
  Mura}, D.~{Landriu}, S.~{Larsson}, L.~{Latronico}, M.~{Lemoine-Goumard},
  J.~{Li}, L.~{Li}, F.~{Longo}, F.~{Loparco}, B.~{Lott}, M.~N. {Lovellette},
  P.~{Lubrano}, G.~M. {Madejski}, F.~{Massaro}, M.~{Mayer}, M.~N. {Mazziotta},
  J.~E. {McEnery}, P.~F. {Michelson}, N.~{Mirabal}, T.~{Mizuno}, A.~A.
  {Moiseev}, M.~{Mongelli}, M.~E. {Monzani}, A.~{Morselli}, I.~V. {Moskalenko},
  S.~{Murgia}, E.~{Nuss}, M.~{Ohno}, T.~{Ohsugi}, N.~{Omodei}, M.~{Orienti},
  E.~{Orlando}, J.~F. {Ormes}, D.~{Paneque}, J.~H. {Panetta}, J.~S. {Perkins},
  M.~{Pesce-Rollins}, F.~{Piron}, G.~{Pivato}, T.~A. {Porter}, J.~L. {Racusin},
  R.~{Rando}, M.~{Razzano}, S.~{Razzaque}, A.~{Reimer}, O.~{Reimer},
  T.~{Reposeur}, L.~S. {Rochester}, R.~W. {Romani}, D.~{Salvetti},
  M.~{S{\'a}nchez-Conde}, P.~M. {Saz Parkinson}, A.~{Schulz}, E.~J. {Siskind},
  D.~A. {Smith}, F.~{Spada}, G.~{Spandre}, P.~{Spinelli}, T.~E. {Stephens},
  A.~W. {Strong}, D.~J. {Suson}, H.~{Takahashi}, T.~{Takahashi}, Y.~{Tanaka},
  J.~G. {Thayer}, J.~B. {Thayer}, D.~J. {Thompson}, L.~{Tibaldo}, O.~{Tibolla},
  D.~F. {Torres}, E.~{Torresi}, G.~{Tosti}, E.~{Troja}, B.~{Van Klaveren},
  G.~{Vianello}, B.~L. {Winer}, K.~S. {Wood}, M.~{Wood}, S.~{Zimmer} and
  {Fermi-LAT Collaboration}, {\it {Fermi Large Area Telescope Third Source
  Catalog}},  {\em \apjs} {\bf 218} (June, 2015) 23
  [\href{http://arXiv.org/abs/1501.02003}{{\tt 1501.02003}}].

\bibitem{icecube_pointsources}
M.~G. {Aartsen}, K.~{Abraham}, M.~{Ackermann}, J.~{Adams}, J.~A. {Aguilar},
  M.~{Ahlers}, M.~{Ahrens}, D.~{Altmann}, K.~{Andeen}, T.~{Anderson} and
  et~al., {\it {All-sky Search for Time-integrated Neutrino Emission from
  Astrophysical Sources with 7 yr of IceCube Data}},  {\em \apj} {\bf 835}
  (Feb., 2017) 151 [\href{http://arXiv.org/abs/1609.04981}{{\tt 1609.04981}}].

\bibitem{egbA}
M.~{Ackermann}, M.~{Ajello}, A.~{Albert}, W.~B. {Atwood}, L.~{Baldini},
  J.~{Ballet}, G.~{Barbiellini}, D.~{Bastieri}, K.~{Bechtol}, R.~{Bellazzini},
  E.~{Bissaldi}, R.~D. {Blandford}, E.~D. {Bloom}, E.~{Bottacini}, T.~J.
  {Brandt}, J.~{Bregeon}, P.~{Bruel}, R.~{Buehler}, S.~{Buson}, G.~A.
  {Caliandro}, R.~A. {Cameron}, M.~{Caragiulo}, P.~A. {Caraveo},
  E.~{Cavazzuti}, C.~{Cecchi}, E.~{Charles}, A.~{Chekhtman}, J.~{Chiang},
  G.~{Chiaro}, S.~{Ciprini}, R.~{Claus}, J.~{Cohen-Tanugi}, J.~{Conrad},
  A.~{Cuoco}, S.~{Cutini}, F.~{D'Ammando}, A.~{de Angelis}, F.~{de Palma},
  C.~D. {Dermer}, S.~W. {Digel}, E.~d.~C.~e. {Silva}, P.~S. {Drell},
  C.~{Favuzzi}, E.~C. {Ferrara}, W.~B. {Focke}, A.~{Franckowiak},
  Y.~{Fukazawa}, S.~{Funk}, P.~{Fusco}, F.~{Gargano}, D.~{Gasparrini},
  S.~{Germani}, N.~{Giglietto}, P.~{Giommi}, F.~{Giordano}, M.~{Giroletti},
  G.~{Godfrey}, G.~A. {Gomez-Vargas}, I.~A. {Grenier}, S.~{Guiriec},
  M.~{Gustafsson}, D.~{Hadasch}, K.~{Hayashi}, E.~{Hays}, J.~W. {Hewitt},
  P.~{Ippoliti}, T.~{Jogler}, G.~{J{\'o}hannesson}, A.~S. {Johnson}, W.~N.
  {Johnson}, T.~{Kamae}, J.~{Kataoka}, J.~{Kn{\"o}dlseder}, M.~{Kuss},
  S.~{Larsson}, L.~{Latronico}, J.~{Li}, L.~{Li}, F.~{Longo}, F.~{Loparco},
  B.~{Lott}, M.~N. {Lovellette}, P.~{Lubrano}, G.~M. {Madejski}, A.~{Manfreda},
  F.~{Massaro}, M.~{Mayer}, M.~N. {Mazziotta}, J.~E. {McEnery}, P.~F.
  {Michelson}, W.~{Mitthumsiri}, T.~{Mizuno}, A.~A. {Moiseev}, M.~E. {Monzani},
  A.~{Morselli}, I.~V. {Moskalenko}, S.~{Murgia}, R.~{Nemmen}, E.~{Nuss},
  T.~{Ohsugi}, N.~{Omodei}, E.~{Orlando}, J.~F. {Ormes}, D.~{Paneque}, J.~H.
  {Panetta}, J.~S. {Perkins}, M.~{Pesce-Rollins}, F.~{Piron}, G.~{Pivato},
  T.~A. {Porter}, S.~{Rain{\`o}}, R.~{Rando}, M.~{Razzano}, S.~{Razzaque},
  A.~{Reimer}, O.~{Reimer}, T.~{Reposeur}, S.~{Ritz}, R.~W. {Romani},
  M.~{S{\'a}nchez-Conde}, M.~{Schaal}, A.~{Schulz}, C.~{Sgr{\`o}}, E.~J.
  {Siskind}, G.~{Spandre}, P.~{Spinelli}, A.~W. {Strong}, D.~J. {Suson},
  H.~{Takahashi}, J.~G. {Thayer}, J.~B. {Thayer}, L.~{Tibaldo}, M.~{Tinivella},
  D.~F. {Torres}, G.~{Tosti}, E.~{Troja}, Y.~{Uchiyama}, G.~{Vianello},
  M.~{Werner}, B.~L. {Winer}, K.~S. {Wood}, M.~{Wood}, G.~{Zaharijas} and
  S.~{Zimmer}, {\it {The Spectrum of Isotropic Diffuse Gamma-Ray Emission
  between 100 MeV and 820 GeV}},  {\em \apj} {\bf 799} (Jan., 2015) 86
  [\href{http://arXiv.org/abs/1410.3696}{{\tt 1410.3696}}].

\end{thebibliography}\endgroup

\end{document}